\documentclass[12pt, oneside]{report} 
\usepackage[utf8]{inputenc}
\usepackage{graphicx}
\usepackage[a4paper, margin=1in, top=2.5cm]{geometry}
\usepackage{setspace}
\usepackage{tocbibind}
\usepackage{tabularx}
\usepackage{multirow}
\usepackage{array}
\usepackage{lipsum}
\usepackage{amsmath, amssymb}
\usepackage{hyperref}
\usepackage{float}
\usepackage{enumitem}
\usepackage{tikz}
\usepackage{pgfplots}
\usepackage{pgf-pie}
\usepackage{smartdiagram}
\usepackage{titlesec}
\usetikzlibrary{positioning, shapes, arrows.meta}
\usepackage[section]{placeins}
\usepackage{longtable}
\usepackage{pdflscape} 
\usepackage{booktabs}
\usepackage{caption}
\usepackage[T1]{fontenc}
\usepackage{mathptmx}

\title{Systematic literature review of the trust reinforcement mechanisms exist in package ecosystems}
\author{Angel Temelko}
\date{\today}

\begin{document}

\begin{titlepage}
    \centering
    \vspace*{2cm}
    {
    {\huge\bfseries Systematic literature review of the trust reinforcement mechanisms exist in package ecosystems}\\
    \vspace{2cm}
    {\Large Angel Temelko, Fang Hou, Siamak Farshidi, and Slinger Jansen}\\
    \vfill
    \hspace*{2cm}\includegraphics[width=0.7\textwidth]{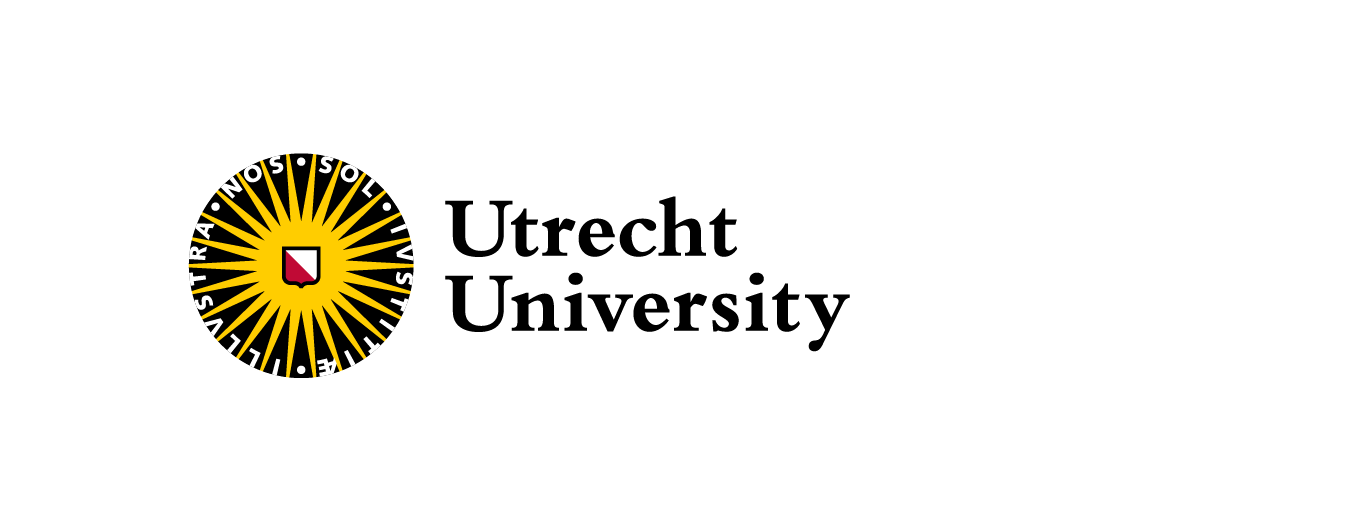}
    \vfill
    Faculty of Science \\
    Master Computing Science \\
    Utrecht University \\
    The Netherlands \\
    {\today}
    }
\end{titlepage}


\newpage

\tableofcontents

\chapter{Introduction}

A "software ecosystem" represents a collaborative and interconnected environment in which a variety of actors, both internal and external to an organization, engage to develop software systems\cite{FormalizingSoftwareEcosystemModeling}.  Historically, companies primarily focused on developing proprietary software for their use. However, recent strategies demonstrate a shift towards greater openness. Today's businesses are actively seeking contributions from broader communities, encouraging them to enhance and refine their software. This approach not only sparks innovation by sharing enterprise software with external groups but also benefits from the expertise these communities bring. This mutually beneficial relationship offers shared advantages~\cite{SofwtareEcoSystemVittorio,KonstantinosSoftwareEcosystemsSystematicLiteratureReview,TrustSECO}.

Open-source software (OSS) systems prioritize code transparency, allowing software stakeholders to access and share it. This model heavily relies on trust, with users expecting the community to provide reliable and safe software~\cite{TrustSECO}. However, there is an inherent risk. In every community, a few might have malicious intentions. When examining platforms such as npm, the network of relationships can be likened to a complicated web. Given this interconnectedness, vulnerabilities in one area can quickly affect many others. For instance, an issue in a single software product can ripple to others that depend on it~\cite{NpmEcoSystemRisk}. It is noteworthy that 40\% of the open source software on npm has known vulnerabilities \cite{NpmEcoSystemRisk}. Similarly, in Java's open-source realm, about one-third are vulnerable \cite{wang2020empirical}.

This raises the question: \textbf{How safe are we within the open-source community?} The answer to this question is complex and multi-faceted, rather than a straightforward affirmation or negation. It is vital that we use software that has undergone checks by trusted organizations, in a manner similar to how Google reviews apps before permitting them on the Play Store. Google Play uses a mix of automated tools and human checks to maintain app and software engineer quality. Google Play Protect (GPP) defends users from possible threats by checking apps, rating their safety, and using ongoing device checks. This involves techniques such as machine learning and both dynamic and static assessments \cite{google_play_protect}. In contrast, package managers, e.g., npm, yarn, lack a strict review process. While people can flag issues, this might not be sufficient \cite{NpmEcoSystemRisk}.

The issues discovered in Log4j clearly showed the security risks tied to open-source software. This flaw allowed harmful strings to be added to the popular Log4j library, enabling unauthorized remote code actions. The so-called Log4Shell problem was not only tough to address but also came with high costs~\cite{hiesgen2022race,log4j1}. Events such as this underline the urgent need for tools that shield both users and software engineers from dangerous open-source software, highlighting the significance of proactive security measures in software development.

Most organizations' codebase is made up of 80\% to 90\% open-source software~\cite{countinThoseThatMatterVulenrab, openSoueceControling}. This underscores the importance of having tools to guard against non-trustworthy open-source software. Regular users, whether they are using a mobile app or an npm library, pose a consistent challenge. They often overlook security warnings, whether on command lines or phones, and do not treat them as urgent.

For instance, a large number of software engineers view updates and vulnerability fixes as extra tasks, not essentials. Surprisingly, 69\% of software engineers are not aware that their codebases have security issues~\cite{kula2018}. Similarly, about 64\% of the general public feels they are not accountable for their smartphone's security~\cite{SecurityAwarnessPhone}.

The remainder of the study is organized as follows. In \textbf{Chapter~\ref{ch:Research Approach}}, we elucidate our research design, elaborating on methodologies such as the Systematic Literature Review, Design Science, and Interview study while also justifying their inclusion. \textbf{Chapter~\ref{ch:LiteratureReview}} reviews existing literature on software trust systems. It contrasts TrustSECO with other notable trust tools in the npm ecosystem, investigates software engineers' security practices and behaviors, and examines their reliance on third-party libraries. This section also outlines the meticulous process of sourcing and categorizing relevant literature, underlining pivotal findings, and spotlighting research gaps. Concluding the research, \textbf{Chapter ~\ref{ch:Conclusion}} reviews the research objectives, summarizes the main conclusions, and recommends directions for future research.

\chapter{Research Approach}
\label{ch:Research Approach}

This research predominantly employs three methodologies: the Systematic Literature Review, Design Science, and Interview study. Each method provides a unique perspective and can address one or more of the research questions formulated. While some questions may benefit from a singular approach, others might require a combination of methods to gain a detailed understanding.

\section{Research Questions}

The primary research question for this study arises from the problem statement:

\begin{itemize}
\item[] \textbf{MRQ:} ``How can package managers reinforce trust within the worldwide Software Ecosystem?''
\end{itemize}

To comprehensively address the MRQ and provide structured insights, it has been subdivided into four additional research questions:

\vspace{1em}

\begin{itemize}
    \item[] \textbf{RQ1:} What kind of trust reinforcement mechanisms exist in package ecosystems?
    \item[] \textbf{RQ2:} How effective are trust reinforcement mechanisms in package ecosystems?
    \item[] \textbf{RQ3:} How do software engineers perceive the balance between adding new features and ensuring security while using third-party packages?
\end{itemize}

\section{Research Methods}

In this study, we conducted a systematic literature review(SLR), design science, and Interview study, which will help us answer our research questions. Every research question can be answered through 1 or more of these methods.

\section{Literature Review}

We conducted a thorough SLR to better grasp the challenges and possible solutions associated with existing npm security tools. Our goal was to delve into documented experiences and findings. Specifically, we were keen to learn about the motivations behind choosing third-party packages, software engineers' responses to warning messages, and their overall understanding of security issues.

The main aim of this review was to pinpoint prevailing trends, methods, and concerns in trust tools for the present npm environment. Furthermore, we sought to understand the complexities of integrating SECO into platforms such as npm. By analyzing earlier studies, our intention was to spot any overlooked areas and steer our research to address them.

A group of researchers explains the nature and purpose of SLRs in the paper titled "Systematic literature reviews in software engineering – A tertiary study" \cite{KITCHENHAM2010792}. According to them, SLRs serve to combine knowledge on a software engineering topic or research query in a manner that is unbiased, transparent, and reproducible. Such reviews are labeled as secondary studies, while the ones they assess are termed primary studies. There are primarily two types of SLRs:
\begin{itemize}
    \item \textbf{Conventional SLRs}
    \item \textbf{Mapping Studies}
\end{itemize}

We are using the systematic review framework outlined by Kitchenham, Barbara in "Procedures for performing systematic reviews"\cite{kitchenham2004procedures}.

\chapter{Literature review}
\label{ch:LiteratureReview}

In this section, we will discuss the outcomes of our thorough examination of previous research, which we conducted using a Systematic Literature Review. Our main goal was to locate significant studies that were relevant to our subject and provide the answers to the initial questions we posed. We looked at all of this research to better understand current tools and their effectiveness. Additionally, we wanted to determine whether software engineers were aware of the problems with these platforms.

\section{Search Strategy}

Our literature search strategy employed a mix of both automatic and manual methods. We initiated our search by identifying relevant keywords that align with our research objectives and the problems we aim to address. The core of our investigation lies in the incorporation of software into package managers, emphasizing the associated challenges and limitations.

Given the multitude of package managers for different programming languages, it was vital to specify our focus. We chose the Node Package Manager (npm) as our primary subject, as that is the tool we aim to integrate with.

 We have also added ``Node Package Manager'' and "Node.js" to widen our search, capturing a more comprehensive view of this ecosystem. By including "software engineer and npm" and "software engineer and packages", we hope to gain insights into software engineers' interactions with npm and packages in general, and also similar tools in the npm ecosystem.

``Best practices'' serve to gather research on industry-recognized standards for npm and building CLI tools. Also to help us retrieve best practices in security for picking third-party packages. 

In the context of npm packages, terms such as ``security'', ``trust systems'', ``vulnerability scanner'', and ``scanner implementation'' are crucial for understanding the security protocols. By examining similar trust systems and vulnerability scanners, we can gain insights into the design and operation of these tools. 

Considering the primary mode of interaction with the npm is through the terminal, ``CLI'' and ``command line interface'' were chosen. These keywords will provide insights into tools, extensions, and best practices software engineers should be aware of when operating within the command-line environment.

The phrase ``open source'' is crucial. It highlights the clear and community-focused approach of many packages in npm. Both TrustSECO's open-source structure and our new tool show the importance of this approach in our study.

``Investigating'', ``Empirical'', ``Characteristics'', and ``Analysis'' are research-oriented terms aiming to narrow down our search to studies or reports that offer in-depth insights and findings on the subject matter.

The word ``survey'' helps us gather opinions, possibly revealing what the larger software engineer community thinks about npm and its use of third-party libraries. This is important for understanding their current needs and issues. It also helps us see how software engineers view security.

Lastly, terms such as "dependencies" and "trivial packages" highlight the details of handling package relationships and the challenges of using packages, whether it is trivial or non-trivial packages\/dependencies.

\noindent \textbf{npm Query:}
\begin{verbatim}
(
    "npm"
    OR "Node Package Manager" 
    OR "Node.js" 
    OR "developer and npm" 
    OR "developer and packages"
)
AND 
(
    "best practices" 
    OR "trust systems" 
    OR "CLI" 
    OR "command line interface"
    OR "integration"
    OR "security" 
    OR "vulnerability scanner"
    OR "scanner implementation"
    OR "open source" 
    OR "Investigating" 
    OR "Empirical"
    OR "Characteristics"
    OR "Analysis" 
    OR "developer" 
    OR "survey" 
    OR "dependencies" 
    OR "trivial packages"
)
\end{verbatim}

We carefully chose the following databases for our systematic literature review:

\begin{itemize}
    \item \textbf{IEEE Explore}
    \item \textbf{Springer}
    \item \textbf{ScienceDirect} 
    \item \textbf{ACM Digital Library}
\end{itemize}

These databases were picked for their trustworthiness, the range of their content, and their relevance to the topic of our study. We aimed to ensure a comprehensive and varied pool of literature for our review by utilizing the advantages of these platforms.

\subsection{Search process}

We obtained a total of 98 articles from the ACM Digital Library, 250 from IEEE Xplore, 374 from ScienceDirect, and 500 from Springer through our automated scholarly search. Thus, a total of 1,222 articles were produced by our primary sources.

\subsection{Duplicate removal}
 
We carried out a duplication removal process based on title and publication year to ensure the originality of the papers acquired and to gain a precise understanding of the volume of our acquisitions. 277 duplicates were found in the primary source dataset, and after removing them, we were left with a total of 945 papers.

\begin{figure}[ht]
    \centering
    \includegraphics[width=0.75\textwidth]{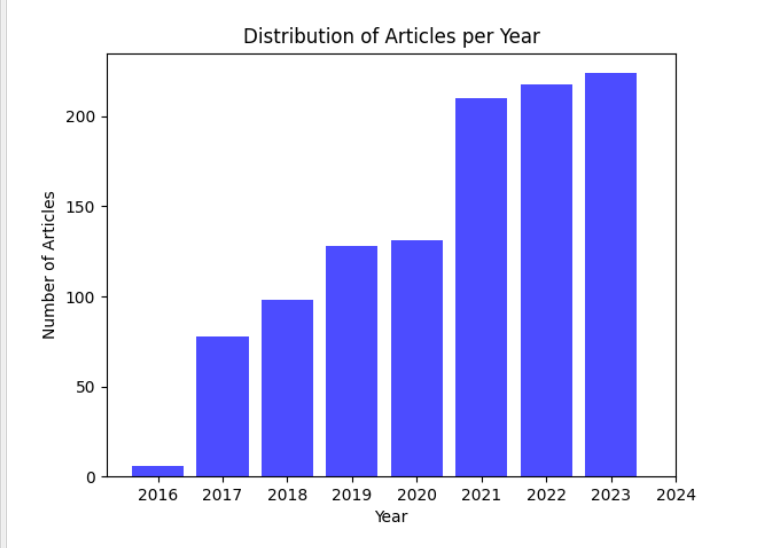}
    \caption{Yearly distribution of selected papers in primary source}
    \label{fig:DistributionOfPapers}
\end{figure}

Interestingly, as shown in Figure \ref{fig:DistributionOfPapers}, there was a marked spike in publications during the years 2021 to 2023. This period seems to have been a breeding ground for discussions related to our topic. This also implies that we looked into the most recent trends in the field.

\section{Inclusion/Exclusion Criteria}

To engage with each paper thoroughly would be neither efficient nor practical given the sizeable volume of more than 945 papers. Establishing strict inclusion and exclusion criteria is essential to speed up the selection process because false positives are a given in any scholarly search.

In the initial round of filtration, a thorough manual review was conducted to narrow down the pool of studies. The criteria for exclusion included:

\begin{itemize}
    \item Papers that were irrelevant or outside the purview of our research goals.
    \item Studies that were books or gray literature.
    \item Studies that were not in English.
\end{itemize}

For each paper, the title and the abstract were carefully examined, and the relevance of each was determined based on its abstract, leading to its inclusion or exclusion accordingly. During this process, a significant number of false positives were identified, particularly within the Springer database. We speculate that these discrepancies arose due to the database's search mechanism and how it queries data. Nevertheless, through meticulous extraction of relevant papers, we managed to refine our list to 147 papers.

\section{Quality Assessment}

We conducted an evaluation of the publications we had incorporated in the next stage of the Systematic Literature Review (SLR). We only used journal articles and conference proceedings as our primary sources, which came from reputable academic libraries. Despite the reputation of our primary libraries, it remains paramount to ensure the quality of each chosen publication. Thus, our evaluation criteria emphasized several key attributes:

\begin{enumerate}
    \item Addressing at least one research question.
    \item Giving a clear statement of the study goal.
    \item Articulating clear and coherent findings.
    \item Presenting a well-defined problem statement.
    \item Focus on Third-party libraries.
    \item Focus on vulnerabilities in npm.
    \item Focus on software engineer security behavior.
\end{enumerate}

After conducting a thorough quality assessment process, which involved a manual review of the papers. We were left with a final count of 57 papers, from which we conducted the process of extracting data and subsequently analyzing and synthesizing the collected information.

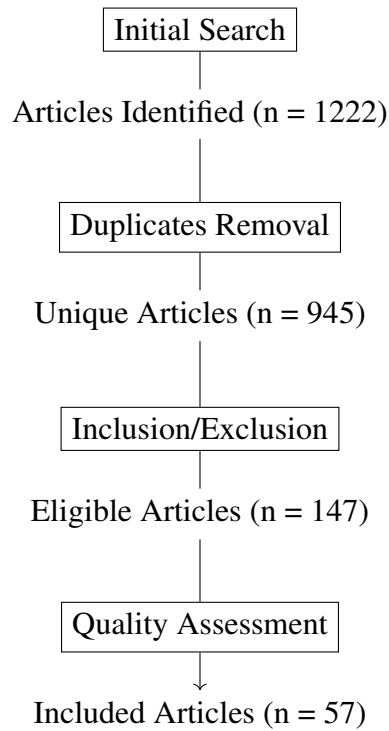
\begin{figure}
    \centering
    \begin{tikzpicture}[node distance=1.5cm, auto]
        \node (initialSearch) [draw, rectangle, align=center] {Initial Search};
        \node (initialRecords) [below=0.5cm of initialSearch] {Articles Identified (n = 1222)};
        \node (duplicatesRemoval) [below of=initialRecords, draw, rectangle, align=center] {Duplicates Removal};
        \node (uniqueRecords) [below=0.5cm of duplicatesRemoval] {Unique Articles (n = 945)};
        \node (inclusionExclusion) [below of=uniqueRecords, draw, rectangle, align=center] {Inclusion/Exclusion};
        \node (eligibleRecords) [below=0.5cm of inclusionExclusion] {Eligible Articles (n = 147)};
        \node (qualityAssessment) [below of=eligibleRecords, draw, rectangle, align=center] {Quality Assessment};
        \node (finalRecords) [below=0.5cm of qualityAssessment] {Included Articles (n = 57)};
        
        \draw[->] (initialSearch) -- (initialRecords) -- (duplicatesRemoval) -- (uniqueRecords) -- (inclusionExclusion) -- (eligibleRecords) -- (qualityAssessment) -- (finalRecords);
    \end{tikzpicture}
    \caption{Flow Diagram representing the article selection process for the Systematic Literature Review.}
    \label{fig:selectionProcess}
\end{figure}

The whole process of our search strategy can be seen in Figure \ref{fig:selectionProcess}

\section{Data Extraction}

After completing our quality assessment, we dived into the papers to extract crucial information. This extraction procedure helped us answer some of our research questions. During this extraction, we focused primarily on two areas: Insight into Tools for Vulnerabilities and Trust Checking in the npm Ecosystem and software engineer Behavior and Dependency on Third-Party Libraries.

\textbf{Insight into Tools for Vulnerabilities and Trust Checking in the npm Ecosystem:} This served as the foundation for RQ1 and RQ2 responses. Our investigation aimed to comprehend the reasons behind and goals of these studies—what issues were they trying to address? We looked at their methodology choices to see if they used exploratory case studies, empirical studies, or other techniques. To gain a thorough understanding of the focus of these studies, a closer examination of the specific features they measured was also conducted. The focus then turned to their measurement metrics: Were they looking at code injections, transitive vulnerabilities, malicious packages, policy validation, or some other features? We tried to understand how their tools worked. To do this, we identified the methods they used, such as machine learning, static analysis, dynamic analysis, or some other technique. To assess the robustness and effectiveness of these tools, we also extracted their effectiveness metrics and results. Additionally, we made note of studies that explicitly stated their limitations because we thought it was important for our research. Finding out whether tools operated before or after installation—i.e., whether the package was evaluated before or after integration—was a crucial point of investigation. All of the data extracted can be seen in Table \ref{table:summaryExistinTools}

\begin{table}[ht]
    \centering
    \begin{tabularx}{\textwidth}{|l|X|}
    \hline
    \textbf{Extraction} & \textbf{Description} \\
    \hline
    Title & The title of the paper \\ 
    Year & The year of the publication \\ 
    Authors & The authors of the paper \\ 
    Source & Which library was selected from \\ 
    Keywords & Keywords of the publication \\ 
    Goal of the study &   What was the goal of the paper \\
    Research method & The research method of the paper \\ 
    Names of the tools & Name of the tool \\ 
    Tool Description & How is the tool described \\ 
    Tool purpose & What is the purpose of the tool \\ 
    Features they measure & Specific features they measured \\ 
    Measurement &  Measurement metrics \\
    Tool type/Approach & What approach they used for the tool \\ 
    Effectiveness Metrics & How effective is the tool \\ 
    Results &  Results that are not contained in the Effectiveness metrics \\ 
    Limitations &  Limitations of the tool \\
    Pre-install / Post-install & Is the tool checking packages before or after installation \\ 
    \hline
    \end{tabularx}
    \caption{Summary of Extracted Data for existing tools}
    \label{table:summaryExistinTools}
\end{table}

\textbf{Software engineer Behavior and Dependency on Third-Party Libraries:} 
The second round of our investigation focused on how software engineers deal with security issues and how much they rely on third-party libraries. We looked at 11 studies on software engineer behaviour and reliance on outside libraries. Understanding the research methods used was a crucial component of our review since it gave us insights into the procedures used to examine software engineer behaviours and third-party library dependencies in light of the chosen study goals. The software engineer's behaviour and Third-party libraries extracted data reliance can be seen in Table \ref{tab:behavioursoftware engineer} and Table \ref{tab:summaryTrivialPackages} respectively.

 \begin{table}[ht]
    \centering
    \begin{tabularx}{\textwidth}{|l|X|}
    \hline
    \textbf{Attribute} & \textbf{Description} \\
    \hline
    Title & The title of the reviewed document or article. \\
    RQ & The research questions posed in the document. \\
    Year & The year in which the document was published. \\
    Authors & The authors of the document. \\
    Source & The source where the document was published. \\
    Keywords & Keywords associated with the document or study. \\
    Goal of the study & The main objective or purpose behind the study. \\
    Methodology & The methodology used in the research or study. \\
    Software engineers' Perceptions & Perceptions and practices of Software engineers as found in the study. \\
    Actual Security Outcomes & The real security results or outcomes as reported in the study. \\
    Factors Influencing software engineers & Factors that influence the software engineers' decisions or behaviors. \\
    Recommendations/Best Practices & Recommended best practices or suggestions made in the document. \\
    Most Common Perception  & The most prevalent perceptions among software engineers as reported in the study. \\
    \% of software engineers (if available) & Percentage of software engineers that hold a certain perception or follow a practice. \\
    Most Adopted Practice  & The most commonly adopted practices by software engineers. \\
    Key Awareness Factor(s) & Major factors or events that raise awareness among software engineers. \\
    \hline
    \end{tabularx}
    \caption{Summary of Extracted Data for software engineers}
    \label{tab:behavioursoftware engineer}
\end{table}

\begin{table}[ht]
    \centering
    \begin{tabularx}{\textwidth}{|p{0.4\textwidth}|X|}
    \hline
    \textbf{Attribute} & \textbf{Description} \\
    \hline
    Document Title & Title of the article. \\
    RQ & Research questions posed. \\
    Year & Publication year. \\
    Authors & Authors of the articles. \\
    Source & Publication source or journal. \\
    Keywords & Associated keywords. \\
    Goal of the study & Main objective of the articles. \\
    Methodology & Research methodology used. \\
    Reasons for Using Packages & Reasons software engineers use such packages. \\
    Empirical Data on Packages & Data derived from analyzing packages. \\
    Consequences of Using Packages & What are the consequences of using such packages \\
    Packages Studied & Total count of packages. \\
    Packages with Tests & Count of packages that have tests. \\
    Packages with High Dependency& packages with a significant dependency count. \\
    Reasons for Using  Packages & Listed reasons for package adoption. \\
    software engineers' Awareness Level & Awareness level of using packages. \\
    software engineers' Adoption Rate & package adoption among software engineers. \\
     packages in open source & Details packages in open-source. \\
    Reasons for Avoiding Packages & Reasons software engineers avoid certain packages. \\
    Recommendation Tools & Indication of tool requirements to recommend packages. \\
    \hline
    \end{tabularx}
    \caption{Summary of Extracted Data on (Trivial) Packages}
    \label{tab:summaryTrivialPackages}
\end{table}

\section{Results}

Upon completing the data extraction, we started by analyzing the data and answering our research questions. The outcomes have been categorized into three significant sections: Trust Reinforcement Mechanisms in npm Ecosystems, software engineers' security practices and behaviours, and utilization of third-party libraries. In the subsequent subsections, a thorough discussion on each of these categories will be conducted separately.

\subsection{Trust Reinforcement Mechanisms in npm Ecosystems}
\label{subsec:npmsecurtitytools}

\begin{table}[ht]
    \centering
    \begin{tabular}{@{}lr@{}}
        \hline
        \textbf{Methodology} & \textbf{Occurrences} \\
        \hline
        Empirical Study & 16 \\
        Design science & 4 \\
        Case study & 2 \\
        Exploratory study & 1 \\
        Corpus analysis / Tool evaluation & 1 \\
        \hline
    \end{tabular}
    \caption{Occurrences of Research Methods in npm}
    \label{tab:occurancesResearchMethodsnpm}
\end{table}

There are 25 papers discussing tools for npm that can check for security-related features such as injections, vulnerabilities, and malicious code. Key information about these tools is presented in Table~\ref{tab:bigtablenpm}. To keep the analysis concise, only the essential fields have been included, while other fields, such as keywords, have been left out to avoid over-complicating the process. The majority of the papers, 16 to be exact, were empirical studies, as can be seen in Table \ref{tab:occurancesResearchMethodsnpm}. As shown in Table \ref{table:toolType}, most of them employed approaches for static and dynamic analysis to check for vulnerabilities. The tools primarily measured vulnerabilities, including taint-style, taint-flow, injection, and security issues. Others, such as Latch\cite{wolf}, took a policy-based approach and prevented users from installing particular packages based on their policies instead of looking for vulnerabilities. Latch is a system designed to mitigate risks associated with install-time software supply chain attacks in the npm ecosystem. Latch primarily focuses on mediating the install-time capabilities of npm packages through an innovative permission system. The tool flagged 100\% of the tested malicious packages and maintainer policies, it is interesting that latch is a pre-installation tool, meaning that it will catch the problem before the user can install the application. 82\% of tested potentially unwanted packages, 100\% of tested malicious packages, and 1.5\% of all npm packages are blocked by it.

\begin{table}[ht]
    \centering
    \begin{tabular}{@{}lr@{}}
    \hline
        \textbf{Tool Type / Approach} & \textbf{Occurrences} \\
    \hline
        Static analysis & 10 \\
        Dynamic analysis & 6 \\
        Taint analysis & 2 \\
        ML & 2 \\
        Anomaly detection & 1 \\
        Query (GitHub Advisory Database) & 1 \\
        Build automation & 1 \\
        Data analysis and modeling & 1 \\
        Benchmark suite & 1 \\
        CLI (not specifically mentioned) & 1 \\
        Analytical tool & 1 \\
        Pattern recognition & 1 \\
        Hashing and content comparison & 1 \\
        Graph-based approach & 1 \\
        Blockchain and smart contracts (Ethereum) & 1 \\
        Code-centric approach (not specifically mentioned) & 1 \\
    \hline
    \end{tabular}
    \caption{Occurrences of Tool Types and Approaches}
    \label{table:toolType}
\end{table}

Two further tools that are associated with permission systems are Demo\cite{demo} and Lightweight Permission System\cite{ligthinhPermissionSystem}. The Demo tool is employed to identify zero-day vulnerabilities inside third-party libraries that are not detectable by npm audit and Snyk. This is achieved by acquiring proper authorization and doing scans on the libraries subsequent to their installation. Static and dynamic analysis are employed to ascertain the necessary permissions for the seamless functioning of these libraries. The study highlights that the exclusive reliance on established tools such as Snyk test and npm audit may overlook possible vulnerabilities, particularly those that are not yet identified. The efficacy of their technology surpasses that of both Snyk and npm. The Lightweight Permission System is designed to provide sandboxes for node.js packages, namely those that do simple calculations, in order to prevent them from accessing security-sensitive resources. The tool has a dual purpose, functioning as both a pre-installation and post-installation mechanism. During the pre-installation phase, software engineers declare the rights required by their packages. Subsequently, the system enforces these permissions to guarantee that packages remain within their designated boundaries. Furthermore, installation prompts serve as reminders to software engineers regarding the declared permissions. The permission system that has been presented has the potential to significantly decrease the amount of work needed to review updates in the analyzed scenario. This reduction in effort can range from 6\% to 52\%.

Visualization tools, such as V-Achilies \cite{vachilies}, leverage the GitHub advisory database to generate graphical visualizations of transitive dependencies. This means the tool not only visualizes the immediate dependencies of a project but also the dependencies of those dependencies. Notably, V-Achilies functions as a post-installation tool. The tool successfully detected vulnerabilities in four of the ten most highly regarded npm projects, namely sinopia, cnpmjs.org, windows-build-tools, and npx. The vulnerabilities exhibited a range of severity and were categorized based on their direct or transitive nature.

The npm-miner\cite{npm-miner} is a tool that has similarities with V-Achilies, although it differs in its absence of visualization capabilities. The primary approach employed by this application for the crawling and analysis of npm JavaScript packages is static analysis. It is important to highlight that the primary data source utilized by the system is GitHub. In the present context, it is noteworthy to include npm-filter\cite{npm-filter} as an additional tool of significance. While it utilizes GitHub data in a similar manner, its primary objective diverges as it focuses on extracting dynamic metadata from program execution. This includes characteristics, namely testing mechanisms, code coverage analysis, and performance metrics. npm-filter is a build automation tool that facilitates the installation, building, and testing of applications within a Docker\footnote{\href{https://www.docker.com/}{https://www.docker.com/}} environment. This tool ensures a regulated and standardized setting for these tasks. It is noteworthy that npm-filter operates as a pre-installation tool, whereas npm-miner operates within a post-installation context.

Another static analysis tool is OpenSSF\cite{openssf} seeks to enhance the security of open-source software (OSS). It functions as an automated tool specifically developed to examine the security status of software packages. OpenSSF utilizes GitHub metrics to evaluate the health of a package before installation. This evaluation is based on various factors such as code reviews, vulnerabilities, licenses, and other features offered by GitHub. The evaluation conducted by the Scorecard tool provided valuable insights into the security procedures throughout the NPM ecosystem. In particular, the documentation of licenses in their respective repositories was found to be present in just 68\% of npm packages. In relation to permissions, the Token-Permission metric revealed that 84.4\% of npm packages exhibited optimum file permissions inside their GitHub processes. Nevertheless, there were also apprehensions. According to the findings of the Scorecard, it was determined that around 15.6\% of npm repositories were found to have \textit{yaml} files that possessed write access rights. This discovery highlights a possible security vulnerability that may be exploited. Moreover, it is worth noting that a significant proportion of npm packages, specifically 30\%, lacked a legal license inside their respective GitHub repositories. In relation to the implementation of security measures, it was found that a significant proportion of npm packages, namely 69\%, did not consistently adhere to Code-Review procedures. Furthermore, a substantial majority of 86\% of these packages showed that they were not subject to regular maintenance. Furthermore, it was observed that around 90\% of npm packages exhibited a lack of implementation of default Branch-Protection and Security-Policy standards inside their repositories. The Scorecard tool utilizes recognized security metrics to assess the security posture of repositories within the npm ecosystem.

Within the domain of static analysis tools, LastJSMile\cite{LastJSmile}, which draws inspiration from LastPyMile\cite{lastPyMile}, has been developed to find code injections in malicious npm packages by discerning inconsistencies between the source code and the package. It has been observed that this tool exhibits a performance improvement of 20.7 times compared to the git-log technique. In the case of authentic npm packages, the tool's regulations resulted in several incorrect notifications. The following static analysis tool is FAST \cite{fast}, designed to identify taint-style vulnerabilities in JavaScript packages, with a focus on achieving a balance between analysis scalability and accuracy. The tool identified a total of 182 zero-day vulnerabilities out of the 242 that were examined. The FAST tool had the most favourable rates of false positives (7.2\%) and false negatives (5.1\%) compared to the other tools that were evaluated.DAPP\cite{Dapp} is designed to automatically identify prototype pollution vulnerabilities inside Node.js modules. Additionally, it is capable of doing parallel analysis on all npm modules. The DAPP system conducted tests on around 75,000 modules, accounting for three-quarters of the total 100,000 modules. The tests revealed an error rate of approximately 26\%. Each module was checked by DAPP in around 6 seconds. Nodest\cite{Nodest} is a feedback-driven static taint analysis tool that is designed to detect injection vulnerabilities in Node.js applications. It employs a feedback-driven approach to do static analysis and identify potential injection vulnerabilities inside the codebase. Nodest demonstrated a commendable level of precision by successfully identifying vulnerabilities in 22 out of 25 modules without any instances of false positives.  The Jam\cite{Jam} tool is utilized to generate accurate call graphs for Node.js in order to gain insights into potential security issues. The precision of Jam is found to be around 84.35\% on average, while \textit{js-callgraph} has an average precision of 58.64\%. The memory rate of Jam is 98.62\%, but the recall rate of \textit{js-callgraph} is 48.16\%. The call graphs generated by Jam exhibited a higher degree of accuracy compared to those created by \textit{js-callgraph}. The duration of Jam's analysis varied, with \textit{toucht} taking less than one second and \textit{jwtnoneify} taking around 23 seconds, both \textit{toucht} and \textit{jwtnoneify} are nodejs packages on which tests were conducted. On the other hand, the performance of js-callgraph was significantly lower. The use of the modular method by Jam resulted in expedited analysis for all benchmarks, with completion times of less than one second seen in several instances. All of these tools employed static analysis techniques at some stage and were categorized as post-installation tools.

During the literature study, two Machine Learning tools were identified in our research. The tool developed by K. Garrett et al~\cite{DetectingSuspiciousPackageUpdates}.  This study presents a novel tool designed to identify malicious package updates within the npm ecosystem. The tool utilizes an analysis of security-relevant features to distinguish between benign and malicious updates. When subjected to testing using recent package updates, the detection model exhibited a significant 89\% reduction in the need for manual review. This promising outcome suggests that the tool has the potential to streamline the verification process for software engineers by effectively flagging suspicious updates. Furthermore, the Amalfi\cite{sejfia2022practical} system incorporates classifiers, reproducibility checks, and clone detection techniques, enhancing its accuracy during the process of retraining. The results of the 10-fold cross-validation conducted on the basic corpus indicated that all of the models exhibited a notably high level of accuracy. However, the recall of the Naive Bayes and Support Vector Machine (SVM) classifiers was lower. This can be attributed to the fact that the dataset had a naturally low proportion of dangerous packages.

We already outlined the tools, Lightweight permission system\cite{ligthinhPermissionSystem}, Demo\cite{demo}, and Latch\cite{wolf} functions in the context of dynamic analysis. We also have three more dynamic analysis tools. The first is 
NodeMedic\cite{NodeMedic} is a software tool that optimizes the operational procedures of the Node.js ecosystem, namely in the areas of triage, vulnerability verification, and the development of package drivers. NodeMedic is a software solution that specifically addresses the server-side dataflow vulnerabilities known as ACE (Asynchronous Code Execution) and ACI (Asynchronous Code Injection). The program revealed vulnerabilities in 40 out of 200 packages, representing a 20\% occurrence rate. Additionally, 85\% of the reported vulnerabilities were effectively exploited with tailored exploits. The second tool in question is NodeXP\cite{NodeXP}. The main objective of the Python program is to automatically identify detection and explanations of Server-Side JavaScript Injection (SSJI) vulnerabilities in Node.js web applications, employing obfuscation techniques to bypass filters and defensive mechanisms. In contrast to other tools, NodeXP was the only tool that successfully identified all the Server-Side JavaScript Injection (SSJI) vulnerabilities present in the apps. The Buildwatch\cite{OhmTowardDetectionOFSupplyChainAttacks}, the last tool utilized in the dynamic analysis approach, effectively identifies and mitigates security vulnerabilities arising from third-party dependencies inside the software supply chain. With the exclusion of buildwatch, a tool that is linked to continuous integration and continuous deployment and might be subject to argument as either a tool used after installation or before installation, the aforementioned dynamic tools are all categorized as post-install tools.

We also discovered benchmarking tools during the data extraction process, such as SecBench.js\cite{SecBench.js}, an executable security benchmark tool for server-side javascript. Contains flaws from advisory databases that cover threat classes, for instance, code injection and path traversal. The benchmark contains 1,244 assertions, averaging 2.07 assertions per exploit.

There are two tools that serve as comprehensive alternatives to npm, effectively replacing or updating the whole npm. The first tool Maxnpm \cite{maxnpm} serves as a substitute for npm and provides a configurable and efficient approach to resolving dependencies. Combines multiple objectives during installation, such as minimizing vulnerabilities and code size.  The MAXnpm tool exhibits enhanced efficacy compared to the conventional npm approach, particularly in its ability to mitigate vulnerable dependencies by 30.51\% and provide newer package solutions with an average improvement of 2.62\%.  The second tool is by D'mello et al. \cite{ethereumBlochain}. The solution makes use of smart contracts and Ethereum's blockchain technology. The technology decentralizes the administration of software packages. In addition to providing an immutable, transparent trace of software provenance, it seeks to avoid potential vulnerabilities discovered in centralized systems. Solidity-based smart contracts are used to manage and store data. They operate in a decentralized environment, with package uploads taking place on peer-to-peer storage. The tool is intended to replace conventional package management systems completely.

In addition, we encountered analytical tools such as DepReveal\cite{DepReveal}, which were designed to examine and comprehend the effects of dependency vulnerabilities in Node.js applications. The fundamental objective of these tools was to enhance software engineers' consciousness about security vulnerabilities present in their dependencies. Out of a total of 200 packages, the system successfully found 40 packages (20\%) that included vulnerabilities. Additionally, it was able to efficiently create exploits for 85\% of the identified vulnerabilities. In addition, we discovered pattern recognition tools, such as the Vulnerability Detection Framework\cite{VulnerabilityDetectionFramework}, which was employed to identify instances of prototype pollution and ReDoS. The precision rate for detecting prototype pollution was 92\%, while for ReDoS it reached 97\%. 

Affogato \cite{AFFOGATO} is a runtime Detection of Injection Attacks for Node.Js using dynamic grey-box taint analysis, it detects all vulnerable flows, meaning it has high recall, and produces no spurious flows, which translates to high precision. Specifically, it detected all vulnerable flows in the given benchmarks and produced no false positives. Poster \cite{poster} tool for detecting malicious code injections in software packages by comparing package repositories with source code repositories using hashing and content comparison approach. The poster needed 12 seconds for processing a source code repository, 0.04 seconds for scanning a suspected artifact, and 33 seconds median execution time for processing the source code repositories, with a 97\% accuracy.

The Unwrapper\cite{unwrapper} tool is a unique solution designed to tackle the problem of "shrinkwrapped clones" within the npm ecosystem. It effectively detects and identifies instances of shrinkwrapped clones, while also identifying similar npm packages that have been cloned. This tool is highly valuable as it helps to address the potential risks associated with package vulnerabilities. By identifying cloned packages, it becomes possible to recognize instances where vulnerabilities are replicated, thereby emphasizing the importance of utilizing such a tool. The precision of the Clone Detector was determined to be 94\%, indicating that out of 100 samples, 94 were correctly identified as genuine positives while six were falsely identified as positives. Additionally, the recall of the Clone Detector was found to be 95.3\%.

Plumber \cite{plumber} is not a regular vulnerability checker, it focuses on the delay in propagation vulnerability fixes in the npm ecosystem, analysis of dependence structures, and package metadata. It provides repair tactics for vulnerable packages. The PLUMBER's efficacy is determined to be 79.8\%, indicating that the proposed repair solutions provided by the tool line up with the actions taken by software engineers in the real-world context in over 80\% of cases.

The last two tools that we are going to check in this Literature review are the DTReme \cite{dtreme} a graph-based approach tool and an extension of the Eclipse-Steady \cite{Eclipse-Steady} tool which uses a code-center approach. DTreme deals with the npm ecosystem's vulnerabilities caused by third-party library dependencies. It seeks to correct errors in current approaches that do not take into consideration npm-specific dependency resolution criteria. The DVGraph insights serve as the foundation for the dependency tree-based vulnerability mitigation tool known as DTReme for npm packages. DTReme outperformed npm's official audit fix in handling vulnerabilities in 77 out of 262 projects.

The Eclipse-Steady\cite{Eclipse-Steady} tool, initially developed for Java and Python programs, was expanded in this study to provide support for JavaScript. The primary objective of this tool is to detect and analyze open-source vulnerabilities inside Node.js applications by employing a comprehensive and code-focused methodology. The identification of occurrences when vulnerable code is repackaged or reused within Node.js applications constituted a fundamental aspect of the tool.

The landscape of tools designed to address npm package vulnerabilities reveals a notable trend. As delineated in Table\ref{table:tolI;nstall}, the majority of these tools operate at a post-installation stage. Only a limited subset is designed for pre-installation checks, while a few ambitious solutions advocate for a comprehensive replacement of npm as the primary package management system.

\begin{table}[ht]
    \centering
    \begin{tabular}{@{}lr@{}}
        \hline
        \textbf{Stage} & \textbf{Occurrences} \\
        \hline
        Post-install & 20 \\
        Pre-install & 4 \\
        Total replacement & 2 \\
        Pre-install / Post-install & 1 \\
        Pre-install on CI/CD & 1 \\
        \hline
    \end{tabular}
    \caption{Occurrences of Installation Stages}
    \label{table:tolI;nstall}
\end{table}

Concerning the metrics used by these tools, there is a pronounced emphasis on assessing various forms of vulnerabilities. Prominently, vulnerabilities related to taint flows, transitive dependencies, and injection attacks are frequently measured. Beyond vulnerabilities, a minority of tools also focus on evaluating the quality and health of packages. Other areas of measurement encompass policy automation, dependency management, and avant-garde techniques and solutions. A comprehensive breakdown of these metrics is detailed in Table \ref{table:groupedMeasurments}.

The reproducability excel for the Table \ref{tab:bigtablenpm} can be accessed at this Excel \href{https://solisservices-my.sharepoint.com/:x:/g/personal/a_temelko_students_uu_nl/EYHptS7bCO9BmPtcbVnmk8ABGaHH7662jvt4mSilqODGFg?e=o6TBwi}{link}

\begin{landscape}
\footnotesize
\captionsetup{font=normalsize}
\setlength\LTleft{0pt}
\setlength\LTright{0pt}
\begin{longtable}
{@{\extracolsep{\fill}}|p{1.8cm}|p{3.3cm}|p{3.0cm}|p{3.0cm}|p{3.3cm}|p{3.3cm}|p{3.3cm}|p{1.8cm}|@{}}

\caption{Summary of the key information of Reinforcement Mechanisms/Tools adopted in npm security checking} \label{tab:bigtablenpm}\\
\toprule
\textbf{Name} & \textbf{Description} & \textbf{Measurement} & \textbf{Type/Approach} & \textbf{Results} & \textbf{Effectiveness Metrics} & \textbf{Limitations} & \textbf{Setup} \\
\midrule
\endfirsthead
\multicolumn{8}{l}%
{{\bfseries Table \thetable\ continued from previous page}} \\
\toprule
\textbf{Tool Name} & \textbf{Description} & \textbf{Measurement} & \textbf{Type/Approach} & \textbf{Results} & \textbf{Effectiveness Metrics} & \textbf{Limitations} & \textbf{Setup} \\
\midrule
\endhead
\bottomrule
\endfoot
\bottomrule
\endlastfoot
NAN  & Detect malicious package updates in Node.js/npm ecosystem. & Malicious package & Anomaly detection / ML & Demonstrated 89\% reduction in manual review effort; identified eslint-scope attack. & 89\% reduction & preliminary evaluation, missing suspicious packages, untested scalability, no explanation mechanism, and no real-time dashboard for suspicious packages. & Post-install \\\hline

V-Achilles & Detect vulnerabilities in npm projects, direct and transitive dependencies, introduces graph UI. & Transitive vulernabilities & vulnerability  & Identified vulnerabilities in 4 top npm projects; direct and transitive dependencies. & NAN & NAN & Post-install \\\hline

Latch  & Tackle risks of install-time software supply chain compromise with Latch (Lightweight instAll-Time CHecker). & Policy vialation & Dynamic analysis & Blocks 1.5\% npm, 82\% undesirable, 100\% malicious packages; 1.6\% workflow impact. &  100\% of tested malicious packages & limitations in portability across operating systems, potential gaps for sophisticated adversaries. & Pre-install \\\hline

npm-miner & Analyze npm JavaScript packages, data management layer, worker processes, and web application components. & Quality of its packages based on maintainability and security & Static analysis & Analyzed 2,000 GitHub packages; found 476K errors, 279K warnings. & NAN & NAN & Post-install \\\hline

npm-filter & Automatically install, build, and test npm packages. & Automation &  build automation & Running isolated in Docker can install, build, and test any application & NAN & Supports only GitHub packages. & Pre-install  \\\hline

OpenSSF & Automate monitoring of open-source software security health. & Health and security package & static analysis and automation tool. & Scorecard: 68\% had licenses, 84.4\% optimal permissions; some practices lacking. & NAN & Focuses on GitHub metrics, potential false positives for non-GitHub projects, does not account for empty repositories. & Pre-Install \\\hline

Plumber & Identify and remedy software vulnerabilities in package ecosystems. & Vulnerabilities within packages,   &  data analysis & PLUMBER: 47.4\% positive feedback from major npm projects. &  nearly 80\%  & Applicability limited to npm packages, results may not generalize to other ecosystems, potential omissions in the dataset, manual inspection errors. & Post-install \\\hline

NodeMedic & NAN & Vulnerabilities, Efficacy of Exploit Synstudy & Dynamic Analysis & NODEMEDIC: Detected vulnerabilities in 20\% of 200 packages, 85\% exploitable. & 85\% of the identified vulnerabilities. & focuses on ACE and ACI vulnerabilities, does not address others, limitations in automation and exploit synstudy. & Post-install \\\hline

SecBench.js & SecBench.js benchmark suite with 600 JavaScript vulnerabilities across major threat classes. & Vulnerabilities & Benchmark suit & SECBENCH.JS: 1,244 assertions, average 2.07 assertions per exploit. & Exploit success via Oracle, 1,244 assertions, avg 2.07 per exploit. & NAN & post-install \\\hline

Lightweight permission system & It is tailored for Node.js applications to minimize attack surface. & Permission system for sandboxes individual packages & dynamic analysis & Permission system reduces update review effort by 6\%-52\%. & 6\% to 52\% reduction. & NAN & Pre-install / Post-install \\\hline

Amalfi & Detect malicious npm packages automatically, addressing vast numbers and updates. & Malicious packages & ML & Models showed high precision; Naive Bayes and SVM had lower recall. & High precision. Low recall &  biases in training data affect generalizability, sustainability issues with continuous retraining, and accuracy issues due to short inspection windows. & Post-install \\\hline

Maxnpm & A complete, drop-in replacement for npm using PACSOLVE and Max-SMT solver. & vulenrabilities & Basic CLI - not mentioned & MAXnpm: Reduces vulnerable dependencies by 30.51\%, slight solving time increase. & 33\% better audit, 2,62\% new packages, 4,37\% code reduction, 1.9\% less duplication. Worse in tests and slower. & The selection may not represent the entire npm ecosystem, possible bugs in tool results, and evaluation criteria may not align with developer priorities. & Pre-install total replacement \\\hline

Nodest & A feedback-driven static taint analysis tool for detecting injection vulnerabilities in Node.js apps. & zero-day injection vulnerabilities, taint flows & Static analysis using feedback-driven approach, taint analysis & Nodest: Detected vulnerabilities in 22/25 modules, no false positives. &  find vulnerabilities in 22 out of 25 modules & Use GitHub only. & Post-install\\\hline

DepReveal & A Node.js project analytical tool for dependency discoverability levels on GitHub. & vulnerabilities  & Analytical tool & DepReveal analyzed dependency vulnerabilities, and provided insights. & Vulnerabilities detected: 40 out of 200 packages (20\%) Exploits successfully crafted: 85\% of the identified vulnerabilities. & Semgrep rules tailored to prototype pollution and ReDoS, may overlook other vulnerabilities, and struggle with obfuscated JavaScript code. & Post-install \\\hline

Vulnerability detection framework & Experimental vulnerability detection framework with Semgrep and textual similarity methods. & vulnerable functions & Pater recognition/ Static analyisis & Detected 18K prototype pollution, 1.7K ReDoS; 92-97\% precision. & Prototype pollution: 92\% ReDoS: 97\%. The precision rate is 98\%. Vulnerable functions detected 94,5\% & NAN & Post install \\\hline

Affogato & A dynamic grey-box taint analysis tool combining black-box and white-box analysis. & injection vulnerabilities & taint analysis tool & Affogato: High precision/recall detecting injection vulnerabilities. &   high recall,  high precision).  & NAN & Post-install \\\hline

Demo & Conduct static and dynamic program analysis on server-side JavaScript third-party libraries. & NAN & Static/Dynamic analysis & Tool surpasses snyk test and npm audit in uncovering unknown vulnerabilities. & More effective than synk and audit & NAN & Post-install \\\hline

Poster  & Compare distributed software artifacts code with source code repositories. & code injection & hashing and content comparison approach & 34 malicious artifacts with code discrepancies; 97\% accurate to source. & 12 seconds for processing a source code repository.0.04 seconds for scanning a suspected artifact.33 seconds median execution time for processing the source code repositories. & NAN & Post-install \\\hline

DAPP & An automatic static analysis tool for prototype pollution vulnerabilities in Node.js modules. & prototype pollution vulnerability & static analysis & DAPP found 37 genuine prototype pollution vulnerabilities in 30K modules. & DAPP: 37 true positives, 38 false positives; 25.68\% error rate; 6.17s avg. test time. & Significant rate of false positives and false negatives, results require manual verification. & Post-install \\\hline

NodeXp & A Python tool for detecting and exploiting SSJI vulnerabilities in Node.js applications. & The Server Side JavaScript Injection (SSJI) vulnerabilities in  & Dynamic Analysis & NodeXP discovered a 0-day vulnerability in SubleasingUIU app. & compared to other tools only NodeXP detected all the SSJI vulnerabilities in the applications. & NAN & post-install \\\hline

DTReme Algortihm  & A dependency tree-based vulnerability remediation tool for npm packages. & vulnerabilities in dependency trees & Graph-based approach (namely terms Graph Tree and DVGraph). & DTResolver 90.58\% accurate in dependency resolution; outperforms npm-remote-ls. &  vulnerability detection, DTResolver achieved coverage of 98.1\% while npm-remote-ls had 97.7\%. & Overlooks indirect dependency vulnerabilities, CVE mapping errors, missing dependencies, and high-risk version exclusions & post-install \\\hline

Eclipse-Steady & Detect, assess, and mitigate vulnerabilities in open source dependencies. & vulnerabilities & code-centric approach - Not mentioned & NAN & Analysis covered 42 out of 65 apps; application constructs outnumber dependency constructs by a factor of three (75 vs. 26). & NAN & Post-install \\\hline

NAN & Manage package interactions via smart contracts, with decentralized verification and a tree data structure. & decentralizing package management using blockchain technology and smart contracts. & Ethereum's blockchain and smart contracts & NAN & Metadata latency: ~148 ms; Peak bandwidth: 650.48 Kbit/s; Other metadata: 396.32 Kbit/s. & Low gas can delay transactions; worker crashes from fund shortages; single failure point; slow node sync; unstable dynamic structure support. & Total replacement \\\hline

LastJSMile & Detect code injections in npm packages by comparing source code with packaged versions, similar to LastPyMile for Python & injections in malicious npm packages & Static analysis & New approach is 89.4\% faster than git-log; reduces false positives. & Tool speed: 20.7x faster than git-log; False positives: 90.2\% for whole artifacts, 21.3\% for "phantom" files; Recall: consistent. & Imbalances in the dataset, dependency on GitHub repos, changes in file hashes from code movement, and overlooking direct malicious code commits. & Post-install \\\hline

FAST & Analyze JavaScript using unique abstract interpretation techniques for efficient vulnerability assessment. & Zero day vulenrabilities, taint-style vulnerabilities & Static analysis / taint style & AST detected 242 zero-day vulnerabilities; 21 CVEs issued. & 182 out of 242 zero-day vulnerabilities. FAST-det had the lowest FP(7.2\%) and FN(5.1\%) & Balancing JavaScript's dynamic features with analytical scalability and accuracy is a key challenge, often leading to a trade-off between the two. & post-install \\\hline

Jam & Build call graphs for JavaScript modules through summary composition. & Security vulnerabilities. & Static analysis & Jam's precision 84.35\%, faster analysis than js-callgraph. & found all 8 vulnerabilities, yielding a 100\% recall. reduced false positives by 81\% compared to NPM audit. The precision is 61\% compared to the 24\% precision of NPM audit.  & NAN & Post-install \\\hline

BuildWatch & Analyze software dependencies dynamically. & Malicious packages,  & Dynamic analysis & NAN & NAN & NAN & pre-install on CI/CD \\\hline

Unwrapper & Detect duplicate NPM packages with a focus on speed and independence from the NPM database. & Clone packages, vulnerabilities in clone packages & file tree structures and file content comparisons. & 10.4\% of 6,000 NPM packages were clones; potential 178K cloned. & Clone Detector's precision is 94\% (94 true positives out of 100 samples, with 6 false positives). Prefilter's recall is 95.3\%. & NAN & post-install \\\hline
\end{longtable}
\end{landscape}

\begin{table}[ht]
\centering
\begin{tabular}{@{}lr@{}}
\hline
\textbf{Measurement Category} & \textbf{Occurrences} \\
\hline
\textbf{Vulnerability Types } &  20\\
- Malicious package & 3 \\
- Transitive vulnerabilities & 1 \\
- Vulnerabilities within packages & 1 \\
- zero-day injection vulnerabilities & 1 \\
- Injection vulnerabilities & 3 \\
- taint flows & 1 \\
- vulnerable functions & 1 \\
- prototype pollution vulnerability & 1 \\
- Server Side JavaScript Injection (SSJI) & 1 \\
- taint-style vulnerabilities & 1 \\
- security vulnerabilities & 1 \\
- injections in malicious NPM packages & 1 \\
- Impact of a vulnerability & 1 \\
\hline
\textbf{Package Quality and Health } & 2 \\
- Quality of packages & 1 \\
- Health and security package & 1 \\
\hline
\textbf{Policy and Automation } & 2\\
- Policy violation & 1 \\
- Automation & 1 \\
\hline
\textbf{Dependency Management } &  3\\
- Dependency relationships & 1 \\
- Evolution of dependencies & 1 \\
- vulnerabilities in dependency trees & 1 \\
\hline
\textbf{Advanced Techniques and Solutions} & 3 \\
- Efficacy of Exploit Synstudy & 1 \\
- Permission system for sandboxes & 1 \\
- Decentralizing package management (blockchain) & 1 \\
\hline
\end{tabular}
\caption{Summary of Measurement Categories Assessed by Tools for Trust Reinforcement in the npm Ecosystem}
\label{table:groupedMeasurments}
\end{table}

\subsection{Software engineers security practices and behavior}
\label{subsec:software engineersSec}

We have analyzed five papers, to understand the security best practices and how software engineers behave towards those securities and practices. 3 papers were Empirical studies, and two papers were Qualitative studies, as can be seen in Table \ref{tab:developerBehResearchMethod}. The data that we extracted can be seen in Table \ref{tab:behaviourDeveloper}

\begin{table}[ht]
    \centering
    \begin{tabularx}{\textwidth}{|l|X|}
    \hline
    \textbf{Attribute} & \textbf{Description} \\
    \hline
    Title & The title of the reviewed document or article. \\
    RQ & The research questions posed in the document. \\
    Year & The year in which the document was published. \\
    Authors & The authors of the document. \\
    Source & The source or journal where the document was published. \\
    Keywords & Keywords associated with the document or study. \\
    Goal of the study & The main objective or purpose behind the study. \\
    Methodology & The methodology used in the research or study. \\
    software engineers' Perceptions & Perceptions and practices of software engineers as found in the study. \\
    Actual Security Outcomes & The real security results or outcomes as reported in the study. \\
    Factors Influencing software engineers & Factors that influence the software engineers' decisions or behaviors. \\
    Recommendations/Best Practices & Recommended best practices or suggestions made in the document. \\
    Most Common Perception  & The most prevalent perceptions among software engineers as reported in the study. \\
    \% of software engineers (if available) & Percentage of software engineers that hold a certain perception or follow a practice. \\
    Most Adopted Practice  & The most commonly adopted practices by software engineers. \\
    Key Awareness Factor(s) & Major factors or events that raise awareness among software engineers. \\
    \% Aware (if available) & Percentage of software engineers who are aware of a certain factor or practice. \\
    \% Unaware (if available) & Percentage of software engineers who are not aware of a certain factor or practice. \\
    \hline
    \end{tabularx}
    \caption{Summary of Extracted Data for software engineers}
    \label{tab:behaviourDeveloper}
\end{table}

\begin{table}[ht]
    \centering
    \begin{tabular}{lr}
        \hline
        \textbf{Methodology} & \textbf{Occurrences} \\
        \hline
        Empirical Study & 3 \\
        Qualitative Study & 2 \\
        \hline
    \end{tabular}
    \caption{Occurrences of Research Methods in software engineer Behavior}
    \label{tab:developerBehResearchMethod}
\end{table}

 Kabir et al. \cite{Security-Relevant-Best-Practices} conducted a study examining three optimal methodologies. In the first step, the user should employ the command "npm audit" to identify vulnerabilities present in library dependencies. Subsequently, these vulnerabilities can be addressed by utilizing the command "npm audit fix". Moving on to the second step, the user should conduct a thorough examination of the packages to identify any unused or duplicated ones. This can be accomplished by employing tools such as "depcheck". Once identified, the user should proceed to delete any redundant packages using the command "npm dedupe". Lastly, to ensure stability and consistency in library dependency versions, it is recommended to enforce the usage of the package-lock.json file. This file serves as a lock file and effectively pins the versions of library dependencies. 
It was discovered that a majority of software engineers do not adhere to best practice BP1, as evidenced by the identification of vulnerabilities in 55\% of the projects examined through the utilization of npm audit. In the context of BP2, it has been observed that a significant proportion of projects have a worrisome abundance of duplicate instances. In the context of BP3, it was discovered that a mere 32\% of the apps surveyed used the practice of explicitly specifying version numbers for their package dependencies. The researchers also aimed to investigate the underlying causes of the violation of these best practices. Their findings indicate that software engineers recognize the significance of security, yet they express scepticism towards npm-audit due to its high rate of false positives. Additionally, it was shown that software engineers often disregarded or misinterpreted alerts regarding duplicate dependencies. A considerable number of engineers did not assign significant importance to concerns related to duplicate dependencies. Certain engineers underlined the necessity of preserving distinct versions of packages, whilst others expressed doubts over the reliability of depcheck. Many software engineers failed to acknowledge the significance of lock files in ensuring consistent builds, either due to a lack of understanding of the functioning of the locking mechanism or due to misunderstandings around it.

The study conducted by Zahan et al~\cite{Security-Practices-Fewer-Vulnerabilities}. aimed to investigate the potential correlation between software security measures and the occurrence of vulnerabilities. The researchers utilized data obtained from the OpenSSF tool \cite{openssf} in order to investigate the correlation between security practices and the number of vulnerabilities. The researchers incorporated a set of 15 established security practices identified in a prior study. Additionally, they collected data from a substantial number of packages, namely 767,389 npm packages and 191,158 PyPI packages. The vulnerability count was obtained from the OSV and Snyk databases. The researchers discovered that the practices of Security Policy, Maintenance, Code Review, and Branch Protection were identified as the most crucial measures for reducing vulnerabilities. It was discovered that packages possessing enhanced security measures frequently exhibited a higher frequency of reported concerns. This phenomenon may be attributed to the higher frequency of usage and more rigorous testing of these programs. Nevertheless, it remains uncertain if the enhancement of security measures directly leads to the discovery of more vulnerabilities, or if there are other contributing elements at play. Additional investigation is required to provide elucidation on this matter.

In a study conducted by Paschenko et al. \cite{Security-Implications}, 25 software engineers were interviewed in a semi-structured manner to gain insight into their decision-making processes regarding the selection, management, and updating of software dependencies. From the interviews, they found out that software engineers often select third-party packages based on company policies, community support, and the library's core functionality, most interesting finding was that software engineers were mostly always focusing on functionality over security. This means a library's ability to perform its intended function is often valued more than its potential security vulnerabilities. When they updated the dependencies, their motivations were, security concerns, they were updating to mitigate vulnerabilities. However, an interesting fact is that software engineers were choosing stability over security, they were only updating it if they knew that it would not crash the application. They were also avoiding dependency updates due to fear of breaking changes. Organizational policies have a substantial impact on whether software engineers update dependencies, and lastly, software engineers find managing and updating dependencies challenging due to the high number of transitive dependencies, which are often difficult to control. They also discussed mitigating unfixed vulnerabilities, they found out that the first thing a software engineer does when faced with a weak dependency is assess the impact it will have on their project. While they wait for an official patch, some people might choose to disable the compromised functionality temporarily. When a vulnerability arises, knowledgeable people frequently take matters into their own hands, addressing the issue on their own and occasionally adding fixes to the original open-source repositories. It may be practical to switch to another library that offers comparable features in situations where patching the vulnerability is complicated or where the affected library is undersupported.

In their empirical study, Kula et al.~\cite{update-dependencies} sought to investigate the behavior of software engineers in updating their library dependencies and their response to security advisories. This investigation was conducted through the examination of 4659 projects and the running of a software engineer survey. The researchers discovered that a significant proportion of software engineers are neglecting to update their library requirements. Specifically, of the projects examined, a staggering 81.5\% were found to have outdated dependencies. A significant proportion, namely 69\%, of the software engineers who participated in the survey shown a lack of awareness of the vulnerabilities present in their software. However, when being notified about these vulnerabilities, they promptly undertook efforts to correct them.  The process of updating these libraries is not simple. Such updates are a difficult choice due to the intricate web of library relationships, also known as "dependency hell." The costs and advantages of updating are frequently compared. These updates are often viewed by software engineers as extra work that is best completed when they have free time. The decision to update also depends on the software engineer's workload and how much weight their team or organization gives to such updates. Even when there are newer alternatives, software engineers favor more established, older libraries. The potential impact of the vulnerability and the role of the dependency in the project are frequently taken into consideration when deciding whether to react to a security advisory. It was claimed by some software engineers that the failure to update the vulnerable dependency was due to the inactivity of the project or due to its lack of criticality. Others believed that the vulnerable element's influence on the project's goals was minimal.

In many instances, software engineers tend to emphasize the usefulness of software over its security, especially when security measures have the potential to impact the product's operational capabilities. Ivory et al.~\cite{roar} found that professionals tend to prioritize feature completion above security due to their demanding schedules and imminent deadlines. It has been shown that software engineers exhibit a prevalent optimism bias, wherein they possess a belief that they can efficiently identify and rectify security vulnerabilities. Indeed, a significant number of security vulnerabilities are typically overlooked. It is noteworthy that certain engineers acknowledge a lack of interest in or familiarity with security, hence rendering their programs more susceptible to vulnerabilities. Moreover, software engineers often rely on their existing expertise, implying that they may only possess an awareness of issues they have already experienced. A significant number of individuals choose heuristic-based coding, a method that may exhibit biases and inaccuracies. They often rely on their friends for guidance instead of consulting security experts, exposing themselves to the potential of receiving inaccurate information. Moreover, certain software engineers may opt to disregard a vulnerability if it poses a risk to the overall operation. Based on the findings of the study, it can be seen that in the absence of a specific motivation, software engineers commonly resort to employing heuristic methods as their default strategy. In many cases, prioritizing functionality is given greater importance than the abstract concept of security due to its tangible nature. Solo software engineers may exhibit a higher degree of idealism and prejudice as a result of limited exposure to contrasting viewpoints. Nevertheless, the prioritization of security by software engineers may be influenced by a company's attention to this aspect.

\subsection{Third-party library usage}
\label{subsec:thridparyt}

The papers that we found in our quality assessment round for Trivial Packages and Third-party library usage are 9. All of them are Empirical case studies, and a comprehensive breakdown of the research methods can be seen in Table \ref{tab:occurancesOfThridPartyLibraries}. The data that we extracted can be seen in Table \ref{tab:summaryTrivialPackages}.

\begin{table}[ht]
    \centering
    \begin{tabular}{lr}
        \hline
        \textbf{Methodology} & \textbf{Occurrences} \\
        \hline
        Empirical case study & 8 \\
        Exploratory study + Empirical case study & 1 \\
        \hline
    \end{tabular}
    \caption{Occurrences of Research Methods in Thrid-party libraries}
    \label{tab:occurancesOfThridPartyLibraries}
\end{table}

The initial study conducted by Abdalkareem R. et al.~\cite{Abdalkareem2017why} aimed to gain insights into the utilization of trivial packages within Node.js applications. The researchers examined a total of 230,000 NPM packages and administered surveys to Node.js software engineers in order to gather information regarding the usage of third-party libraries and the concept of "trivial packages".Upon conducting an investigation, it was determined that a package is classified as trivial if it contains fewer than 35 lines of code and had a McCabe's cyclomatic complexity of less than 10. More than 50\% of the trivial packages examined lacked test coverage, had fewer releases, and 43.7\% had at least one dependency. Moving to the software engineers, the researchers discovered that the utilization of third-party libraries and trivial packages is mostly driven by their well-implemented and well-tested nature. These tools contribute to enhanced productivity, maintainability of code, improved readability, and reduced complexity within the application. There were instances in which individuals claimed that it enhanced their performance. It was determined that the software engineer community has a significant level of awareness on the benefits and challenges associated with utilizing trivial packages and third-party libraries.

Abdaklerem R, et al. \cite{pythonTrivial} extended their previous work, included 501,001 packages, and also tested the PyPl together with npm. They found out that trivial package definitions are the same for JavaScript and Python. software engineers of JavaScript and Python see trivial packages differently. In contrast to Python software engineers, who see the usage of trivial packages as a negative practice in 70.3\% of cases, just 23.9\% of JavaScript software engineers thought it was harmful to use such packages. software engineers predominantly use trivial packages for their well-implemented and tested nature, with 54.6\% of JavaScript and 54.1\% of Python respondents citing this reason. 47.7\% of JavaScript and 32.4\% of Python software engineers believe these packages enhance productivity. A minority of respondents, 9.1\% from JavaScript and 5.4\% from Python held the belief that Well-maintained code is a significant factor. They identified several drawbacks, A significant 55.7\% of JavaScript and 67.6\% of Python software engineers face ``dependency hell''. Additionally, there is the risk of application breakage. As noted by 18.2\% of JavaScript and 32.4\% of Python software engineers. Performance can be impacted too, with 15.9\% of JavaScript and 8.1\% of Python software engineers citing slower build, run, and installation times. In some cases, instead of speeding up work, trivial packages can cause delays (12.5\% JavaScript, 10.8\% Python). While 9.1\% of JavaScript software engineers mention potential missed learning opportunities, both communities are particularly concerned about security: 8.0\% of JavaScript and a notable 13.5\% of Python software engineers underscore the vulnerabilities trivial packages might bring.

The paper by M. A. R. Chowdhury et al.~\cite{TrivialPackagesMAR} also explored the reasons for adopting these trivial packages, they found out that most of the trivial packages are deeply integrated into projects, meaning if the packages are down it will affect the project. The overwhelming influence trivial packages had on both individual projects and the larger npm ecosystem surprised many software engineers. The information sparked reflection, and several software engineers hypothesized that the prevalence of trivial packages might be due to the void left by the lack of a solid standard library for JavaScript. Another interesting observation was that when considering whether to include dependencies, software engineers frequently gave priority to elements such as community acceptance and active project activity, frequently ignoring the triviality. They stressed that software engineers who use these packages should not undervalue their significance. Before incorporating these packages into their projects, they should make sure they are regularly updated and carefully reviewed. Notably, these unimportant packages continue to be project dependencies after their initial inclusion as well. To reduce dependency risks, software engineers are urged to rigorously assess possible dependencies while taking into account their trivial nature and investigate refactoring or migration approaches.

In their study, Enrique Larios Vargas et al.~\cite{enriqueSelectingLibraries} examined the perspective of software engineers in the process of selecting third-party libraries. A total of 16 software engineers were interviewed, and a survey was conducted with 115 software engineers. The examination encompassed the technical, human, and economic factors that software engineers consider when choosing third-party libraries. 
In relation to the technical aspect, the software engineers expressed their preference for libraries that are active and maintained long-term, they have regular updates, recent releases, and active contributions, and provide insights into library vitality. The quality elements that have an impact on a library include good documentation, usability, alignment with their architecture, good test coverage, no security vulnerabilities, and good performance. The most fundamental criterion was the extent to which the library fulfilled its project requirements. Another noteworthy observation is that software engineers who initiate projects from the beginning have greater autonomy in choosing libraries, but for ongoing projects, library selection is impacted by the necessity to conform with the existing software.

Xu et al \cite{reinventingTheWheel} conducted two surveys exploring library reuse and re-implementation to discern why software engineers either substituted self-created code with an external library or vice versa. Their findings revealed that 69.6\% of respondents agreed on the commonality of replacing self-implemented code with a library method, with 83.9\% admitting to having done this in their practice. Only 39.3\% believe that replacing a library method with a self-implemented code is common, but a larger 76.8\% have done this in their practice. The reasons for replacing the library method they gave were Improving reliability by 25\%, Development efficiency by 24\%, Testability by 22\%, and Maintainability by 20\%. Their criteria for library selection were as follows: 22 out of 34 participants were in Library maintenance and testing, for Library reputation, there were nine participants, Code and documentation readability had four participants, and Stability had four participants. Library size/complexity had three participants, License compatibility had 3 participants and finally, Integration ease had 2 participants. The findings reveal that due to a lack of knowledge of the library or because the library's method was not yet introduced, software engineers frequently replaced their internally implemented methods with those from external libraries. They decide against using their own implementation when they later come across a well-maintained and tried library that meets their needs in the context of why software engineers Replace an External Library Method. With their Self-Implemented Code, they found out that 21\% wanted to reduce dependency, 19\% wanted to improve flexibility, 18\% wanted a simpler solution, and the interesting fact was that only 3\% wanted better security.

Mujahid et al \cite{highly-selected-packages} conducted a qualitative study on a survey of JavaScript software engineers to determine the qualities that JavaScript software engineers want in npm packages. They surveyed 118 JavaScript software engineers for this study. The researchers discovered that the primary factors influencing the choice of a library include documentation, which received a 93\% preference rate, followed by download counts at 85\%. The star count, rated on a 5-point scale, averaged at 4.0. Lastly, around 62\% of software engineers considered vulnerabilities when making their selection. The parameters that exhibited some degree of significance were release dates, commit frequency, test code, license, dependent applications count, number of dependencies, closed issues count, and number of contributors.  Another noteworthy consideration for software engineers when picking libraries was the level of support from prominent companies such as Facebook, Formidable Labs, and Infinite Red. Libraries backed by these companies were perceived as more trustworthy and dependable. Positive reception is often observed for active community conversations and endorsements from reputable software engineers within the community.

In their study, Chen et al.~\cite{helpingnothelping} conducted a survey including 59 Javascript software engineers who actively create trivial npm packages. The objective of the study was to investigate the motivations behind software engineers' decision to publish such packages, as well as their perceptions of the negative repercussions associated with these products. Additionally, the researchers aimed to assess the extent to which these negative concerns may be alleviated. The results obtained from the study indicated that the advantages of publishing simple packages were as follows. The primary focus of this study is on the development of reusable components, which accounts for 64.41\% of the overall emphasis. Additionally, the examination also encompasses the evaluation and documentation of these components, constituting 33.90\% of the research concerns. The concept of separation of concerns, with a significance of 32.2\%, is another crucial aspect that is explored. Furthermore, the study delves into the optimization of these components, which has an importance of 27.12\%. Lastly, the study acknowledges the importance of contributing to the community and personal satisfaction, which accounts for 22.03\%. When they were asked what the disadvantages were, the respondents predominantly indicated the need to manage various packages, the emergence of dependency conflicts, and the challenge of identifying suitable packages. They proposed grouping their packages, and it will save 13\% of the number of dependencies in the ecosystem.

Lopez et al. \cite{fernanndoo1} proposed several metrics for software engineers to consider when selecting libraries:

\begin{itemize}
    \item \textbf{Popularity:} Number of client projects using the library.
    \item \textbf{Release Frequency:} Average interval between consecutive releases.
    \item \textbf{Issue Response And Closing Times:} Average times for issue responses and issue resolutions.
    \item \textbf{Recency:} The date of the latest library update or release.
    \item \textbf{Backwards Compatibility:} Average number of breaking changes per release.
    \item \textbf{Migration:} Frequency of library replacements in client projects.
    \item \textbf{Fault-proneness:} Measured by the number of bug fixes.
    \item \textbf{Performance \& Security:} Efficiency of the library's code and its vulnerability.
\end{itemize}

In a subsequent study, Lopez et al.~\cite{fernandoo} conducted research aimed at examining the efficacy of software metrics in facilitating software engineers' decision-making process when selecting libraries. The majority of participants expressed that metric-based comparisons were beneficial, as shown by an average Likert scale rating of 3.85. Performance, Popularity, and Security were the top three metrics influencing software engineers’ decisions. Their means were 4.08, 4.06, and 4.00, respectively.
Other metrics, such as Issue Closing Time and Last Discussed on Stack Overflow were found to be less influential. The software engineers are in need of supplementary metrics pertaining to the usability of documentation and libraries. The significance of specific indicators differs across different areas. For instance, the importance of security varies across different areas.

\section{Software Ecosystem}
\label{sec:softeco}
Before delving into TrustSECO's background and functionalities, it is essential to understand the concept of the software ecosystem: its definition, the various types of ecosystems, and their mining capabilities. 
SECOs are defined as \textit{"are sets of actors that collaboratively serve a market for software and services, typically with an underlying technical platform"}~ \cite{Hou2022,jansen2013software}. The roles within the software ecosystem include~\cite{Hou2022}:

\begin{itemize}
    \item \textbf{End-user}: An individual who uses the software to improve their productivity.
    \item \textbf{End-user Organization}: A group of end-users using the software to support their organization's objectives.
    \item \textbf{Software Engineer}: Professionals who develop and maintain software products.
    \item \textbf{Software Producing Organization}: An entity that employs software engineers to develop and maintain software aimed at broad adoption.
    \item \textbf{Package Maintainers}: Software engineers responsible for the upkeep of software packages, often working independently and utilizing platforms like GitHub for storage and maintenance.
\end{itemize}

Furthermore, they describe the exchanges within the software ecosystem, which consist of:

\begin{itemize}
    \item \textbf{Software Product}: \textit{"A prepared collection of software elements or a service powered by software, complete with supporting materials, made available for a particular market."}
    \item \textbf{Component}: \textit{"An independent module with specified connectors and requirements that can be assembled and implemented on its own."}
    \item \textbf{Library}: A compilation of functions tailored for specific tasks, such as React, which provides functions that facilitate writing client-side code.
    \item \textbf{Package}: A collection that includes software, libraries, and metadata detailing the name, version, and dependencies of the software.
\end{itemize}

\begin{figure}
    \centering
    \includegraphics[width=1\linewidth]{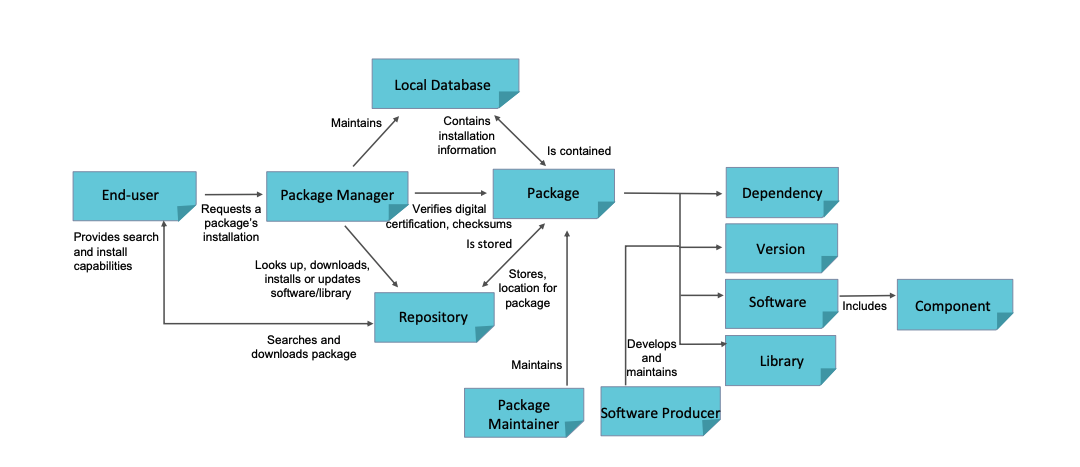}
    \caption{Structure of the software package ecosystem\cite{Hou2022}.}
    \label{fig:sofecofang}
\end{figure}

Lastly, the authors elucidate the following ecosystem services in Figure \ref{fig:sofecofang}.

\begin{itemize}
    \item \textbf{Ecosystem Services}: Services that enhance existing software, such as Integrated Development Environments (IDEs) that assist in code writing and debugging.
    \item \textbf{Package Manager}: A tool for the automated management and updating of software packages on a computer system.
    \item \textbf{Package Repository}: A storage server where software packages are hosted and made accessible.
\end{itemize}

More researchers provide an exploration of the SECO. According to Mens and De Roover, SECO can be viewed from various perspectives, including ecological, economic, technical, and social, each perspective offers a unique context and structure~\cite{MensRoover2023}. This understanding is from Manikas\cite{Manikas2013SoftwareEcosystems} that encapsulates all these views, describing a software ecosystem as: \textit{``the interactions of a set of actors on top of a common technological platform, resulting in a number of software solutions or services. Each actor has specific motivations or business models and is connected to other actors and the ecosystem through symbiotic relationships. The technological platform is structured to allow the involvement and contribution of the different actors''}. Additionally, they further categorize software ecosystems into six primary types:

\begin{itemize}
\item \textbf{Digital Platforms:} These are defined as platforms where the product is owned by a company, such as Apple's App Store, JetBrains' IntelliJ IDEA, or Microsoft's VSCode. They allow third-party apps or plugins to integrate, with contributors being the software engineers of these third-party libraries and their respective users.

\item \textbf{Social Coding Platforms:} These platforms primarily host software projects. They serve as repositories where software engineers can version, store, and maintain their codebases. Examples include GitHub, GitLab, and BitBucket. In this context, the components are the software projects, and the contributors are the software project software engineers.

\item \textbf{Component-based Software Ecosystems:} This concept is centered around the reuse of software components, enabling the efficient utilization of previously developed software, thereby saving time and reducing costs. Initially proposed by McIlroy, this idea did not gain much traction during its inception. However, with the advent of cloud computing and the emergence of OSS, it has witnessed significant success. A prime example of such ecosystems is the realm of third-party libraries, which is the focal point of this study. Each ecosystem is accompanied by its own package managers and package registries where the software packages are stored. For instance, Python utilizes PyPi, JVM languages use Maven, JavaScript employs npm, and .NET relies on NuGet. The core components of these ecosystems are the interdependent software packages, while the contributors encompass both the consumers and producers of software packages and libraries.

\item \textbf{Software Automation Ecosystems:} These ecosystems are centred around automating various facets of software management, development, and deployment. Examples include:
\begin{itemize}
\item \textit{Containerization:} Software engineers package components into containers, ensuring uniform behavior across different environments.
\item \textit{Management:} Tools such as Bicep and Terraform, which use Infrastructure as Code (IaC), help automate infrastructure management tasks.
\item \textit{DevOps and CI/CD:} Continuous deployment and delivery tools streamline and automate workflows during deployments. Here, the components are container images and CI/CD pipelines, and the contributors are DevOps professionals.
\end{itemize}

\item \textbf{Communication-oriented Ecosystems:} Unlike technical ecosystems, these primarily focus on platforms facilitating communication within software communities. Examples include modern communication platforms such as Slack and Discord, and discussion forums, namely Stack Overflow where software engineers seek and offer advice. The components in these ecosystems are emails, messages, posts, and questions, while the contributors encompass software engineers, end-users, and researchers.

\item \textbf{OSS Communities:} OSS communities are decentralized groups that maintain open and transparent projects. While they offer the benefits of transparency and openness, challenges such as delayed updates and unmaintained components persist, often because many contributors volunteer their time unpaid. However, initiatives are emerging to financially support these contributors. Prominent OSS communities include the Apache Software Foundation and the Linux Foundation.
\end{itemize}

The above classification of SECOs reveals their various roles in the industry. Together, these ecosystems form a complex environment that is critical for understanding modern software development, particularly with regard to third-party libraries.

\section{The npm Ecosystem}

In this study, we delve into the realm of Component-based software ecosystems, with a specific focus on the npm ecosystem. This selection is due to a discernible gap in the literature and the intended design of a tool tailored for this ecosystem. Npm, an acronym for Node Package Manager, is a package manager for the Node.js platform. According to the official npm documentation, the current count of packages is 2,565,715.

Package managers, as defined by Hisman et al. \cite{packagemanagers}, are \textit{``programs that map relations between files and packages (which correspond to sets of files), and between packages (dependencies), facilitating users in maintaining their systems at the package level rather than dealing with individual files''}. In essence, package managers simplify the task of packaging files and code, promoting reusability within our codebases. This is the core functionality of npm, which aids developers in packaging reusable code to be used within the Node.js environment \cite{npmDocs}.

npm is recognized as the world's largest software registry. The npm ecosystem comprises three primary components:

\begin{itemize}
\item Website: Utilized for searching and examining packages.
\item CLI (Command Line Interface): Employed by developers to manage packages, including retrieving and uploading them to the registry with commands such as \textbf{npm install} and \textbf{npm publish}.
\item Registry: The database that contains all the JavaScript packages.
\end{itemize}

We can understand the roles within the npm ecosystem by referencing the model presented by Hou and Jansen, as shown in Figure~\ref{fig:sofecofang}. The primary users of npm are software engineers and organizations that utilize the packages available through npm's Command Line Interface (CLI). Npm itself acts as the package manager, overseeing a registry that stores packages and facilitates their retrieval and installation upon user request. Additionally, npm maintains a local database that aids in managing the metadata of installed packages. The packages in the npm registry are maintained and kept up-to-date by software engineers. The package can also have dependencies in the npm, other dependent packages which are called transitive dependencies, the npm automatically detects and installs all the packages.

They also explain the trust factors in the software package managers, the factors that contribute towards the trust are dependency hell, security vulnerability, and package prevention. In npm there is no such solution to dependency hell, when you install a package it will give you all the dependencies, npm is not strong enough with prevention as explained in \cite{NpmEcoSystemRisk}, most of the high vulnerabilities and preventions are done when a user reports it and they can take down the package, also regarding the vulnerabilities, there can be packages with vulnerabilities and the only thing that npm is doing is giving you with warning using npm audit, and an ability to fix with npm audit fix, however, it does not detect all the vulnerabilities and it does not fix all of them.

Compared to other Software Ecosystems (SECOs), npm is distinct as it is open-source, allowing everyone to contribute and view the code. This contrasts with many other SECOs that are closed-source. Another unique aspect of npm is its centralization around a single repository, the npm registry. Governance in npm is community-driven, which sets it apart from others. Additionally, npm excels in dependency management, offering robust features in this area. Furthermore, the ease of versioning and publishing in npm is notable, making it more user-friendly compared to other SECOs.

\section{Background on TrustSECO}

TrustSECO is an innovative system designed to help users confidently choose and install software based on its reliability. Its main spotlight is on tracking and assessing different versions of software packages. The system gathers insights from a wide array of users, including regular software users, software-making companies, and even the individuals behind the creation of these software packages. They all come together to pool information that provides insight into the software's past performance, any known issues, and its general trustworthiness. This gathered data is housed in an online ledger that serves as the foundation of TrustSECO. using a specially created scoring mechanism, this ledger actively assists in determining how much a user can trust a specific software package. By ensuring users have all the information they require about software before deciding to use it, TrustSECO essentially aims to make the digital world safer and more dependable~\cite{TrustSECO}. In the following subsection, we explain and dive into each part of the TrustSECO software.

\subsection{Distributed Ledger}

The Distributed Ledger was designed to decentralize the database, allowing the community to create and access it. The community ensures the system remains private and secure to prevent the entry of unchecked data, employing ``spider jobs'' that come with a variable fee.

This ledger lets both users and software providers input data. Based on the information provided, the ledger adjusts the score: factors such as the number of GitHub stars can boost the score, while the presence of harmful bugs might reduce it.

\subsection{Spider} The spider is in charge of collecting the information or trust facts via API calls or web crawling.  It pulls data from several sources, including Stack Overflow, Libraries.io, GitHub, and CVE to collect the data.

\subsection{Portal} 
The portal of the project is created in Vue.js\footnote{\href{https://vuejs.org/}{https://vuejs.org/}}. The main purpose of the portal is to provide UI for the user, so they can spider and/or retrieve/store data on the distributed database.

\subsection{Trust Score Calculation}
Through the use of a \textit{trust score}, TrustSECO has developed a system to evaluate the trustworthiness of software packages. This rating has several properties:

\begin{enumerate}
     \item \textbf{Multi-dimensionality}: It assesses the software package as well as its particular version and the software engineer community that supports it.
    \item \textbf{Transparency}: The score calculation is replicable given the same data, promoting community consensus.
    \item \textbf{Combinatorial}: The evaluation of package combinations can be done by combining various scores.
    \item \textbf{Numeric}: Designed to quickly determine whether a piece of software is reliable.
\end{enumerate}

According to TrustSECO, software trust has many facets and is influenced by both the software and additional factors. They have a number of properties that they are calculating based on TrustSECO's scoring. Starting with \textbf{Packages + Versions}, because trust is variable, different versions of the same package might have different trust ratings. Bug resolution times, user feedback, known vulnerabilities, and confidence in the contributing software engineers are all factors that affect this. Secondly, the sole \textbf{Package Managers}, they are potentially unsafe. TrustSECO's assessment criteria encompass Usage frequency. Any malicious, outdated, or broken dependencies. Reputation and popularity. Compliance with security standards. Recent compromises of the package manager. Another part of the calculation involves the \textbf{Software Engineers}, who are critical to the collection of the Trust facts because they are providing information such as activity duration on platforms such as Github, star ratings, and negative experience.
Last, it is the \textbf{Software Organizations}, they are considering the following factors: Organizational support and popularity.

The steps for Trust Score calculation are as follows:

\begin{enumerate}
    \item A user submits a package name and version.
    \item Relevant trust facts for that package are retrieved.
    \item Each trust fact's data points are counted and stored in an array.
    \item Each data point, converted to an integer, is multiplied by its weight. This value is then divided by the number of data points for that trust fact to obtain an average.
    \item The final score is adjusted to fit within a 0-100 range
\end{enumerate}

\begin{figure}[ht]
    \centering
    \includegraphics[width=0.8\textwidth]{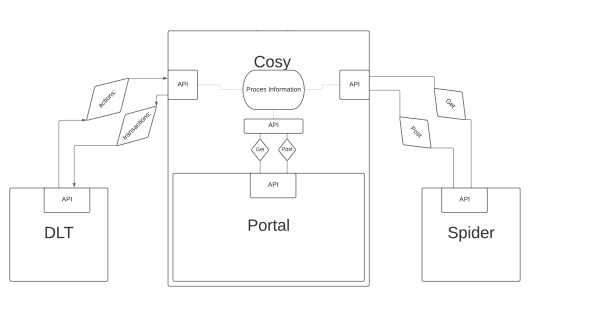}
    \caption{Architecture for Integration of TrustSECO with npm \cite{TrustSECO2022}.}
    \label{fig:ArchitectureTrustSeco}
\end{figure}

The model of the TrustSECO can be seen in Figure \ref{fig:ArchitectureTrustSeco}

The implementation details and source code of TrustSECO are available for review on GitHub at \href{https://github.com/SecureSECO/TrustSECO}{this link}.

\section{Grey Literature Review}
In the systematic literature review (SLR), we identified various tools for trust reinforcement mechanisms within the npm ecosystem. A notable gap was observed in the literature regarding the development of such tools. Given that our tool is intended to be command-line based, a decision influenced by the common interaction of developers with npm through CLI, we explored this direction further. The npm ecosystem primarily comprises three components: the website, CLI, and the registry. While the website facilitates library exploration and the registry serves as a storage medium, the CLI remains the most interactive element for software engineers. Hence, our focus is on developing a CLI wrapper tool over npm. However, existing literature lacks detailed insights into the development methodologies, libraries used, and best practices followed in the creation of such tools. Consequently, we turned to grey literature to bridge this knowledge gap.

For adherence to best practices in CLI tool development, we referred to sources such as Prasad et al. \cite{clig2023}, Jeff D \cite{12FactorCli}, official Heroku documentation \cite{heroku}, Gnu Standards for Command Line Interfaces \cite{gnu-standards-2023}, and Czapski \cite{czapski2022}.

Czapski emphasizes the development of command-line interfaces (CLIs) across various programming languages, highlighting key guidelines for optimal CLI design. These guidelines include ensuring command clarity, producing easily interpretable outputs, and enhancing command discoverability.

Jeff presents a structured approach with 12 critical factors for building effective CLIs that they introduced in Heroku. These encompass the necessity of comprehensive help features, a preference for flags over arguments, inclusion of version control, attention to input and output streams, robust error handling, user-friendly interfaces, interactive prompts, the use of tabular data representations, optimizing for performance, fostering a community for contributions, and clarity in subcommand structures.

The official GNU documentation advocates for simplicity and user-friendliness in CLI design. It recommends the implementation of both short and long-named options, such as '-o' and '--options', to cater to user preferences. A fundamental requirement is the inclusion of '--version' and '--help' flags in all CLIs. Additionally, it stresses the importance of maintaining consistency in commands, flags, and arguments throughout the program.

Prasad's guide offers a thorough framework for creating effective command-line interface (CLI) tools. His methodology focuses on a 'human-first' design, prioritizing user experience in CLI development. He emphasizes the critical role of comprehensive documentation and readily accessible help functions to guide users. The guide also delves into the importance of designing intuitive and clear outputs, ensuring that users can easily understand and interact with the CLI.

Error handling is another key aspect highlighted by Prasad, underscoring the need for CLIs to manage and report errors effectively. He discusses the optimal use of arguments and flags to enhance user command control, as well as the significance of CLI interactivity for a more engaging user experience. Subcommands are addressed, with a focus on their organization and clarity.

Furthermore, Prasad underscores the importance of robustness and future-proofing in CLI tools. This includes considerations for scalability and adaptability in response to evolving user needs and technology advancements. Lastly, he provides insights into effective naming conventions, ensuring that command and option names are intuitive and self-explanatory, thereby enhancing overall usability.
 
Our exploration in the npm registry and various articles led us to identify the following libraries that aid in CLI tool development:

\begin{itemize}
\item \textbf{yargs:} 90,783,468 Weekly Downloads, 7 dependencies, last release: 7 months ago, Health Analysis: 91/100.
\item \textbf{oclif:} 124,840 Weekly Downloads, 18 dependencies, last release: 1 month ago, Health Analysis: 86/100.
\item \textbf{minimist:} Weekly Downloads: 56,176,805, Dependencies: 0, last release: 9 months ago, Health Analysis: 88/100.
\item \textbf{meow:} 18,440,305 Weekly Downloads, Dependencies: 0, last release: 3 months ago, Health Analysis: 91/100.
\item \textbf{inquirer:} Weekly Downloads: 28,800,689, Dependencies: 15, last release: 6 days ago, Health Analysis: 97/100.
\item \textbf{vorpal:} 34,767 Weekly Downloads, Dependencies: 10, last release: unknown, Health Analysis: 53/100.
\item \textbf{commander.js:} Weekly Downloads: 134,435,650, Dependencies: 0, last release: 2 days ago, Health Analysis: 93/100.
\end{itemize}

The Health Analysis is done with Snyk. Among these seven npm libraries, oclif is classified as the most comprehensive framework. However, its popularity is moderate, it has numerous dependencies, and its health rating is below 90. In contrast, yargs, minimist, meow, and commander.js are more lightweight and provide essential features for CLI development. Inquirer is a utility for interactive prompts. Vorpal, however, is no longer supported and has a significantly low health rating.

In the evaluated npm libraries, commander.js stands out as a notable example, striking an optimal balance between being lightweight and feature-rich. It emerges as the most popular among its counterparts, characterized by its extensive maintenance. Remarkably, commander.js operates with zero dependencies and has achieved a health score of 93, further solidifying its position as a preferred choice in this domain.

\section{Discussion}

With the SLR and all the data that we gathered, we can answer our research questions. We have 25 papers to study to answer RQ1 and RQ2, and 15 papers for RQ3, the data extracted can be seen in Table \ref{table:rqpapers}

\begin{table}[ht]
    \centering
    \begin{tabular}{lr}
        \hline
        \textbf{Research questions} & \textbf{Papers} \\
        \hline
        RQ1 & 29 \\
        RQ2 & 29 \\
        RQ3 & 14 \\
        \hline
    \end{tabular}
    \caption{Research questions with number of papers}
    \label{table:rqpapers}
\end{table}

\textbf{RQ1: What kind of trust reinforcement mechanisms exist in package ecosystems?}

 \textbf{Answer:} Addressing this question is complex. Initially, we must define what constitutes trust in the software ecosystem. According to Hou et al.\cite{TrustSECO2022}, it is the end-user's willingness to take risks, grounded in their belief that the system providers will be dependable. Based on this definition, trust in package ecosystems involves the end-user trusting that both the package provider and creator will be reliable. However, trust cannot always be assured, especially in less stringent open-source systems like npm. While npm strives to enforce security and dependability, vulnerabilities can still be introduced. As noted in subsection \ref{subsec:npmsecurtitytools}, there are numerous tools enhancing trustworthiness in these ecosystems. These include various approaches like dynamic and static analysis, build automation, sandboxing, and machine learning techniques to identify vulnerabilities. These tools focus on different aspects, such as transitive vulnerabilities, code injections, zero-day vulnerabilities, clone packages, malicious code, and permission systems, each with its specific use cases. The npm ecosystem performs checks for potential security issues, but some risks may still bypass these measures. Notably, most tools do not prevent the installation of a package but rather operate as post-installation tools, scanning the codebase or requiring manual package checks, as shown in Table \ref{table:tolI;nstall}. We identified a gap in our study: the need for a tool that safeguards users before they install a package, allowing them to assess its trustworthiness.

 \textbf{RQ2: How effective are trust reinforcement mechanisms in package ecosystems?}

 \textbf{Answer:} The evaluation of trust reinforcement mechanisms in package ecosystems, as reported in the literature, generally indicates positive outcomes. Tools analyzed in various studies demonstrated high precision in detecting issues and security vulnerabilities in libraries. However, there's a variance in the reported effectiveness. Some studies did not disclose effectiveness metrics, while those that did revealed differing perceptions of effectiveness. This variation is often attributed to factors like the occurrence of false positives or negatives, performance measurements, and the number of errors identified.

In terms of capabilities, certain tools were effective in identifying basic errors, whereas others were advanced enough to detect zero-day vulnerabilities. The diversity in techniques used by these tools complicates the task of establishing a uniform standard for comparing their effectiveness. Additionally, the literature often points to limitations such as high rates of false positives or negatives, reliance on specific platforms like GitHub, and constraints unique to particular tools. These factors collectively contribute to the challenge of assessing the overall efficacy of trust reinforcement mechanisms in package ecosystems.

 \textbf{RQ3:How do software engineers perceive the balance between adding new features and
ensuring security in their npm packages?}

 \textbf{Answer:} In Subsection \ref{subsec:software engineersSec} and Subsection \ref{subsec:thridparyt}, we examined the literature related to the utilization of third-party libraries and the security awareness of software engineers. Our analysis revealed that software engineers demonstrate an increased level of security consciousness when selecting third-party libraries, regardless of whether they are complex libraries or simple packages. Nevertheless, while selecting libraries or attempting to upgrade them, the emphasis is consistently placed on prioritizing the library's usefulness or the stability of the system rather than its security. Security vulnerabilities are not their first priority. When selecting a library, individuals tend to value indicators such as documentation quality, download numbers, and star counts. A significant proportion of software engineers are unaware of the vulnerabilities present in their libraries, and many of them neglect to update these libraries due to perceiving it as a burdensome task.

Based on our examination of Research Questions 1, 2, and 3, it is apparent that a deficiency exists within the current body of literature. To begin with, it is worth noting that there is currently a lack of available tools that offer proactive warnings or safeguards prior to the installation process. Furthermore, it is evident that software engineers acknowledge the need of security; nevertheless, it is not frequently prioritized within their top three factors when making package selections. It is believed that the implementation of this tool will serve as a preventive measure against the installation of potentially harmful third-party libraries. This is due to the observation that software engineers generally lack the initiative to thoroughly assess security vulnerabilities when selecting packages, as well as the limited effort put into updating existing packages. By "preventing" software engineers from initiating installations altogether, it is anticipated that a significant portion of the challenges currently encountered by software engineers can be mitigated.

\chapter{Conclusion}
\label{ch:Conclusion}

In this study, we conducted a thorough literature review on npm security tools, developer security practices and behavior, and the use of third-party libraries. Our examination of various tools revealed a significant gap: the lack of pre-installation safeguards for npm packages. This was evident when compared to the predominantly post-installation focus of tools such as Plumber \cite{plumber}, Affogato \cite{AFFOGATO}, and NodeXp \cite{NodeXP}, as well as others discussed in section \ref{subsec:npmsecurtitytools}. This finding is particularly crucial considering that developers often prioritize functionality over security \cite{Security-Relevant-Best-Practices, Security-Practices-Fewer-Vulnerabilities, Security-Implications, roar}.

\bibliographystyle{plain}
\bibliography{references}

\end{document}